\documentclass[prb,aps,twocolumn,floats,showpacs,preprintnumbers,amsmath,amssymb]{revtex4}

\usepackage{graphicx}
\usepackage{dcolumn}
\usepackage{bm}

\newcommand{\bq}{\begin{equation}}
\newcommand{\eq}{\end{equation}}
\newcommand{\bqa}{\begin{eqnarray}}
\newcommand{\eqa}{\end{eqnarray}}
\newcommand{\nn}{\nonumber \\}

\def\be     {\begin{equation}}
\def\ee     {\end{equation}}
\def\bea        {\begin{eqnarray}}
\def\eea        {\end{eqnarray}}
\def\bnn    {\begin{eqnarray*}}
\def\enn    {\end{eqnarray*}}

\begin{document}

\title{Vertex correction and Ward identity in the U(1) gauge theory with Fermi surface}
\author{Ki-Seok Kim$^{1,2}$}
\affiliation{ $^1$Asia Pacific Center for Theoretical Physics,
Hogil Kim Memorial building 5th floor, POSTECH, Hyoja-dong, Namgu,
Pohang 790-784, Korea  \\ $^2$Department of Physics, Pohang
University of Science and Technology, Pohang, Gyeongbuk 790-784,
Korea }
\date{\today}

\begin{abstract}
We show that introduction of vertex corrections in the fully
self-consistent ladder approximation does not modify dynamics of
spinons and gauge fluctuations in the U(1) gauge theory with Fermi
surface.
\end{abstract}


\maketitle

\section{Introduction}

Spin liquid has been a cornerstone in the gauge theory approach
for strongly correlated electrons \cite{Lee_Nagaosa_Wen} as Fermi
liquid in the Landau-Ginzburg-Wilson framework for phase
transitions \cite{Shankar}. An effective field theory is often
given by compact U(1) gauge theory \cite{Polyakov}, implying that
"magnetic monopole" excitations should be irrelevant in order to
be self-consistent for the theory. Two kinds of mechanisms have
been proposed, resulting from either spinon dynamics with Fermi
surface \cite{Hermele,SungSik_Deconf,Nagaosa,KS_Deconf} or
existence of a topological term \cite{Senthil} associated with
anomaly in the Dirac theory \cite{Tanaka,KS_ZeroMode}. After
deconfinement is demonstrated, an important task is to solve the
non-compact U(1) gauge theory.

A standard technique is the large-$N$ approximation, where the
spin degeneracy of a spinon is extended from $\sigma = \uparrow,
\downarrow$ to $\sigma = 1, ..., N$. The $N \rightarrow \infty$
limit was believed to suppress higher-order quantum
loop-corrections in the Fermi surface problem
\cite{Rech_Pepin_Chubukov} just as the case of the relativistic
invariant theory \cite{Hermele}. Recently, it was clearly
demonstrated that the Fermi surface problem is still strongly
interacting even in the large-$N$ limit, meaning that all planar
Feynmann diagrams should be summed as the non-abelian gauge theory
with Lorentz invariance \cite{SungSik_Genus}.

This observation suggests that dynamics of fermions (spinons) and
gauge fluctuations can be modified by more-loops quantum
corrections, that is, the exponent in the frequency dependence of
the spinon self-energy may have a nontrivial correction, affecting
transport properties of spinons. Actually, the lowest-order vertex
correction associated with the Aslamasov-Larkin diagram is shown
to cause such a correction proportional to $1/N$ although the
lowest-order vertex correction associated with the ladder diagram
does not change the dynamics of both fermions and collective
bosons \cite{Metlitski_Sachdev}.

In this paper we perform an infinite-order summation for the
ladder-type vertex correction and find no anomalous correction for
the exponent in the frequency dependence of the fermion
self-energy. In other words, dynamics of both spinons and gauge
bosons remains the same as the case without vertex corrections
\cite{McGreevy}. We prove this result based on the Ward identity
\cite{Chubukov}, asymptotically exact in the low energy limit.

\section{Beyond the Eliashberg framework}

\subsection{Review on the Eliashberg theory}

We start from an effective U(1) gauge theory with one patch in one
time and two space dimensions \cite{SungSik_Genus} \bqa && {\cal
L} = f_{\sigma}^{\dagger} \Bigl( \eta
\partial_{\tau} - i
\partial_{x} - \partial_{y}^{2} \Bigr) f_{\sigma}
+ \frac{e}{\sqrt{N}} a f_{\sigma}^{\dagger} f_{\sigma} + a (-
\partial_{y}^{2})^{\frac{z-1}{2}} a , \nn \eqa where $f_{\sigma}$
and $a$ represent fermionic spinons and U(1) gauge fluctuations,
respectively, emerging in the U(1) spin liquid state
\cite{Lee_Nagaosa_Wen}. $e$ is an internal gauge charge of a
spinon and $N$ is the spin degeneracy. $\eta$ is an infinitesimal
coefficient to control artificial divergences in quantum
corrections, which can be cured by self-energy corrections
\cite{SungSik_Genus}. $z$ is the dynamical exponent determining
the dispersion relation of gauge fluctuations. It is given by $z =
3$ for several problems such as ferromagnetic or nematic quantum
criticality including the present spin liquid problem
\cite{Rech_Pepin_Chubukov} while $z = 2$ in the spin density wave
ordering \cite{HMM}. Both the Fermi velocity $v_{F}$ and the
curvature $1/m$ are set to one.

It was shown that either scattering with small momentum transfer
or that with 2$k_{F}$ (twice of the Fermi momentum) is relevant in
the Fermi surface problem \cite{Shankar,SungSik_Deconf}. In the
one patch approximation \cite{SungSik_Genus} only forward
scattering, identified with $g_{4}$ in the $g$-ology of the one
dimensional problem \cite{Maslov_Review}, is taken into account
while another forward scattering ($g_{2}$) and back scattering
($g_{1}$) are neglected. Such scattering channels can be
introduced in the two patch approximation
\cite{Metlitski_Sachdev}.


The previous large-$N$ analysis without vertex corrections
coincides with the Eliashberg approximation
\cite{Rech_Pepin_Chubukov}, introducing only self-energy
corrections \bqa \Pi(q_0,q) & = & \frac{e^{2}}{N} \int \frac{d
k_{0}}{2\pi} \int \frac{d^{2} k}{(2\pi)^{2}}
G_{\sigma}(k_0+q_0,k+q) G_{\sigma}(k_0,k)
\nn & = & \gamma_{b} \frac{|q_{0}|}{|q_{y}|} , \nn \Sigma(k_0) & =
& - \frac{e^{2}}{N} \int \frac{d q_{0}}{2\pi} \int \frac{d^{2}
q}{(2\pi)^{2}} G_{\sigma}(k_0+q_0,k+q) D(q_0,q)
\nn & = & - i \frac{\lambda_{b} }{N} \mbox{sgn}(k_0) |k_0|^{2/z} ,
\eqa where the spinon Green's function and gauge propagator are
given by \bqa && G_{\sigma}(k_0,k) = \frac{1}{i \eta k_{0} + k_{x}
+ k_{y}^{2} - \Sigma(k_0)} , \nn && D(q_0,q) =
\frac{1}{|q_{y}|^{z-1} + \Pi(q_0,q)} , \eqa respectively. The main
point is that dynamics of gauge fluctuations is given by the
Landau damping term with the damping coefficient $\gamma_{b}$
proportional to $k_{F}^{-1}$, resulting from Fermi-surface
fluctuations, while spinon dynamics shows non-Fermi liquid
dependence in frequency of its self-energy, given by $2/z$ with a
constant $\lambda_{b}$.
%
%

The problem to address in this paper is whether the anomalous
exponent $2/z$ will be modified or not when vertex corrections are
taken into account in a non-perturbative way, i.e., up to an
infinite order. It turns out that the gauge dynamics cannot be
modified from the Landau damping dynamics in the one patch
approximation while the fermion dynamics is expected to have some
corrections \cite{SungSik_Genus}. Even in the two patch
approximation, the gauge dynamics is still unchanged while
fermions were shown to have $1/N$ modification for the frequency
exponent in the perturbative approach based on the Eliashberg
solution \cite{Metlitski_Sachdev}.

\subsection{Self-consistent ladder approximation}

We introduce vertex corrections in a completely non-perturbative
way based on the ladder approximation. Then, we obtain full
self-consistent equations \bqa && \Pi(q_0,q) = \frac{e^{2}}{N}
\int \frac{d k_{0}}{2\pi} \int \frac{d^{2} k}{(2\pi)^{2}}
\Lambda(k_0+q_0,k+q;k_0,k) \nn && G_{\sigma}(k_0+q_0,k+q)
G_{\sigma}(k_0,k)
, \nn && \Sigma(k_0) = - \frac{e^{2}}{N} \int \frac{d q_{0}}{2\pi}
\int \frac{d^{2} q}{(2\pi)^{2}} \Lambda(k_0+q_0,k+q;k_0,k) \nn &&
G_{\sigma}(k_0+q_0,k+q) D(q_0,q) ,
\eqa where $\Lambda(k_0+q_0,k+q;k_0,k)$ is the vertex function
given by \bqa && \Lambda(k_0+q_0,k+q;k_0,k) = 1 - \frac{e^{2}}{N}
\int \frac{d l_{0}}{2\pi} \int\frac{d^{2} l}{(2\pi)^{2}} \nn &&
\Lambda(k_0+q_0-l_0,k+q-l;k_0-l_0,k-l) D(l_0,l) \nn &&
G_{\sigma}(k_0+q_0-l_0,k+q-l) G_{\sigma}(k_0-l_0,k-l)  \eqa in the
ladder approximation. Fig. 1 represents Eq. (4) and Fig. 2
displays Eq. (5).

\begin{figure}[t]
\includegraphics[width=0.30\textwidth]{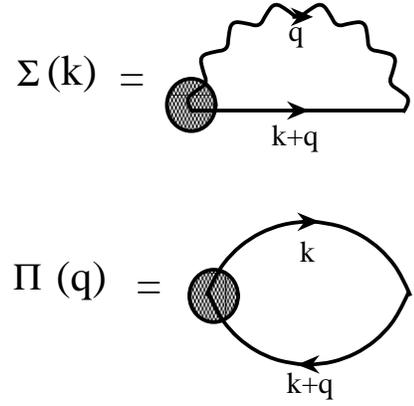}
\caption{Fermion self-energy $\Sigma(k)$ and boson self-energy
$\Pi(q)$, where the thick line represents the fermion Green's
function and the wavy line does the gauge propagator. The shaded
region can be any renormalized vertex.} \label{fig1}
\end{figure}

\begin{figure}[t]
\includegraphics[width=0.35\textwidth]{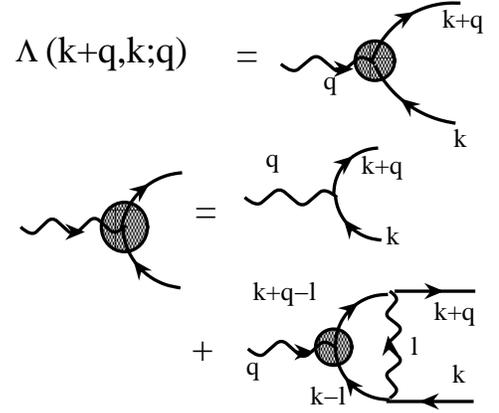}
\caption{(Color online) The ladder vertex correction turns out to
be irrelevant in the Eliashberg solution.} \label{fig2}
\end{figure}

In order to solve these three coupled integral equations, we
consider the Ward identity \cite{Maslov_Review} \bqa && i \eta
q_{0} \daleth (k_0+q_0,k+q;k_0,k) + q_{x}
\Lambda(k_0+q_0,k+q;k_0,k) \nn && + q_{y} \digamma
(k_0+q_0,k+q;k_0,k) \nn && = G_{\sigma}^{-1}(k_0+q_0,k+q) -
G_{\sigma}^{-1}(k_0,k) , \eqa where $\daleth (k_0+q_0,k+q;k_0,k)$
is the scalar vertex while $\Lambda(k_0+q_0,k+q;k_0,k)$ and
$\digamma (k_0+q_0,k+q;k_0,k)$ are vector vertices. There is a
special relation between the scalar vertex and vector one in one
dimension due to the kinematic constraint, that is, the vector
vertex is proportional to the scalar vertex with the Fermi
velocity \cite{Maslov_Review}. This is the reason why the one
dimensional problem is exactly solvable. On the other hand, such a
relation does not exist above one dimension, thus it is necessary
to propose an ansatz for such a relation.

We suggests the following relation \bqa \daleth
(k_0+q_0,k+q;k_0,k) & \rightarrow & \Lambda(k_0+q_0,k+q;k_0,k) ,
\nn \digamma (k_0+q_0,k+q;k_0,k) & \rightarrow & (2k_{y}+q_{y})
\Lambda(k_0+q_0,k+q;k_0,k) , \nn \eqa where the first ansatz is
the application of the one dimensional result while the second one
is our main assumption. As a result, the vertex function is \bqa
&& \Lambda(k_0+q_0,k+q;k_0,k) = \frac{G_{\sigma}^{-1}(k_0+q_0,k+q)
- G_{\sigma}^{-1}(k_0,k)}{g_{\sigma}^{-1}(k_0+q_0,k+q) -
g_{\sigma}^{-1}(k_0,k)} , \nn \eqa where $g_{\sigma}^{-1}(k_0,k) =
i \eta k_0 + k_x + k_y^2$ is a non-interacting Green's function.
We note that this expression recovers both non-interacting and one
dimensional cases. In appendix A we show that this ansatz is
self-consistent for the vertex equation [Eq. (5)] in the low
energy limit.

It is educational to check that if we apply the well-known one
dimensional ansatz for the vertex function, i.e., neglecting
$\digamma (k_0+q_0,k+q;k_0,k)$, given by \bqa &&
\Lambda(k_0+q_0,k+q;k_0,k) \approx
\frac{G_{\sigma}^{-1}(k_0+q_0,k+q) - G_{\sigma}^{-1}(k_0,k)}{i
\eta q_{0} + q_{x} } , \nonumber \eqa we find that the Landau
damping dynamics for gauge fluctuations is not reproduced as
follows \bqa && \Pi(q_0,q) = \frac{e^{2}}{N} \frac{1}{i q_{0} +
q_{x}} \int \frac{d k_{0}}{2\pi} \int \frac{d^{2} k}{(2\pi)^{2}}
\Bigl( G_{\sigma}(k_0,k) \nn && - G_{\sigma}(k_0+q_0,k+q)\Bigr)
\propto e^{2} \frac{i |q_0|}{i \eta q_{0} + q_{x} } . \nonumber
\eqa This implies that the one dimensional Ward identity cannot be
applied to higher dimensional cases.

Inserting the vertex function Eq. (8) into the equation for the
polarization function in Eq. (4), we find that the Landau damping
dynamics does not change as follows \bqa && \Pi(q_0,q) \nn && =
\frac{e^{2}}{N} \int \frac{d k_{0}}{2\pi} \int \frac{d^{2}
k}{(2\pi)^{2}} \frac{G_{\sigma}(k_0,k) -
G_{\sigma}(k_0+q_0,k+q)}{i\eta q_{0} + q_{x} + 2k_y q_y +
q_{y}^{2}} \nn && = \frac{i e^{2}}{2} \int \frac{d k_{0}}{2\pi}
\int \frac{d k_{y}}{2\pi} \frac{\mbox{sgn}(k_0+q_0) -
\mbox{sgn}(k_0)}{i\eta q_{0} + q_{x} + 2k_y q_y + q_{y}^{2}} \nn
&& = \frac{e^{2}}{8} \int \frac{d k_0}{2\pi}
\frac{\mbox{sgn}(q_0)[\mbox{sgn}(k_0+q_0) -
\mbox{sgn}(k_0)]}{|q_y|} \nn && = \gamma \frac{|q_{0}|}{|q_{y}|} ,
\eqa where $\gamma$ is a modified damping coefficient.

Inserting both the vertex function [Eq. (8)] and boson self-energy
[Eq. (9)] into the equation for the fermion self-energy in Eq.
(4), we obtain two sectors \bqa && \Sigma(k_0) = - \frac{e^{2}}{N}
\int \frac{d q_{0}}{2\pi} \int \frac{d^{2} q}{(2\pi)^{2}}
G_{\sigma}(k_0+q_0,k+q) D(q_0,q) \nn &&
\frac{G^{-1}_{\sigma}(k_0+q_0,k+q) - G^{-1}_{\sigma}(k_0,k)}{i\eta
q_{0} + q_{x} + 2k_y q_y + q_{y}^{2} } \nn && \equiv
\Sigma_{1}(k_0) + \Sigma_{2}(k_0) .
%
%
\eqa The first part denoted by $\Sigma_{1}(k_0)$ turns out to
vanish  \bqa && \Sigma_{1}(k_0) = \frac{i e^{2}}{2 N} \int \frac{d
q_{0}}{2\pi} \int \frac{d q_{y}}{2 \pi} \frac{
\mbox{sgn}(q_0)}{\gamma \frac{|q_{0}|}{|q_{y}|} + |q_{y}|^{z-1} }
= 0 . \nn \eqa On the other hand, the second part recovers exactly
the same expression as Eq. (2) in the low energy limit  \bqa &&
\Sigma_{2}(k_0) = \Sigma(k_0) = - i \frac{\lambda }{N}
\mbox{sgn}(k_0) |k_0|^{2/z} , \eqa where $\lambda$ is a modified
constant. We show the derivation of this result in appendix B.

\subsection{Discussion}

An essential point is that the ladder-vertex equation [Eq. (5)] is
solved, resorting to the Ward identity, where the ansatz for the
relationship between the scalar and vector vertices [Eq. (7)] is
introduced to result in the relationship between the vertex
function and fermion Green's function [Eq. (8)]. Justification for
Eq. (8) lies in the fact that the boson self-energy should be
given by the Landau damping solution. In the one patch formulation
higher order quantum corrections are shown to vanish identically
because all poles in the integral expression are in the same half
plane, implying that the Landau damping solution is exact
\cite{SungSik_Genus}. Considering the structure of the boson
self-energy with the ladder-vertex correction [Eq. (9)], Eq. (8)
seems to be generic. In addition, Eq. (8) recovers not only the
non-interacting case but also the one dimensional physics
completely. The relationship between Eq. (7) and Eq. (8) is unique
as far as the y-current vertex is linearly related with the
x-current or scalar vertex.

Can we use the Ward identity in this approximation scheme? Usually
speaking, the Ward identity is on the relationship between full
vertex corrections and corresponding Green's functions. Actually,
the Ward identity and the special relation between the vector and
scalar vertices in one dimension are satisfied for the fully
renormalized vertex. In fact, the Ward identity should be always
satisfied in any approximation scheme because it guarantees
conservation of the system. In this sense the Ward identity may be
regarded as another phrase of the conserving approximation.
Mathematically speaking, the conserving approximation can be
derived from the Luttinger-Ward functional approach, where fully
self-consistent sets of equations derived from the Luttinger-Ward
functional respect the Ward identity automatically \cite{LW1,LW2}.
Of course, this is not exact. A good example can be found in the
impurity problem, called the conserving self-consistent t-matrix
approximation \cite{LW_Impurity}.

Another important assumption is that singular dependence of the
fermion self-energy occurs from frequency instead of momentum.
Though this is a common result within the Eliashberg approximation
\cite{Rech_Pepin_Chubukov}, there is no reason a priori, for this
assumption to remain valid as soon as ladder-type vertex
corrections are included. In particular, the self-consistent
calculations performed in the appendices are carried out within
this assumption, which greatly simplifies the computations. It was
argued that the fermion self-energy has the same frequency
dependence as the Eliashberg solution and there is no singular
momentum dependence in the perturbative evaluation of the one
patch formulation up to an infinite order based on the Eliashberg
solution \cite{SungSik_Genus,McGreevy}, implying no anomalous
exponent for the fermion Green's function, although one can not
remove the possibility that the summation for coefficients from
higher order quantum corrections may be singular. In addition,
ladder-type vertex corrections turn out not to change the
Elaishberg solution in the perturbative calculation of the two
patch formulation up to the lowest order \cite{Metlitski_Sachdev}.
In this respect our result is not surprising but expected from the
perturbative analysis in both one-patch \cite{SungSik_Genus} and
two-patch \cite{Metlitski_Sachdev} formulations.

However, special types of quantum corrections involved with
$2k_{F}$ momentum transfer were shown to cause the singular
momentum dependence for the fermion self-energy in the two patch
formulation, giving rise to an anomalous exponent for the fermion
Green's function \cite{Metlitski_Sachdev}. Unfortunately, these
quantum processes are not introduced in the ladder approximation,
given by the Aslamasov-Larkin diagrams. This leads us to consider
the Aslamasov-Larkin vertex up to an infinite order, shown in Fig.
3. Frankly speaking, this consideration is not completely new,
already investigated in the context of the superconducting
instability although it corresponds to the particle-particle
channel \cite{BCS}. An interesting point is that such vertices are
singulary enhanced in the U(1) spin liquid state, causing
anomalous critical exponents according to the perturbative
evaluation. There is another Aslamasov-Larkin vertex correction in
the particle-particle channel associated with superconductivity,
competing with the $2k_{F}$ particle-hole instability. One problem
in this consideration is to construct the self-consistent
conserving approximation, not addressed clearly as far as we know.
When this self-consistent conserving framework is settled, we can
check whether the vertex function [Eq. (8)] from the Ward identity
allows the self-consistent solution with new critical exponents or
not. If it works, we have a powerful framework.

\begin{figure}[t]
\includegraphics[width=0.45\textwidth]{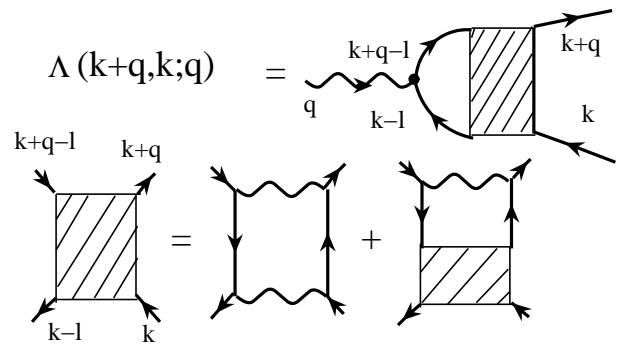}
\caption{(Color online) The Aslamasov-Larkin vertex correction in
the particle-hole channel, where $2k_{F}$ momentum transfer gives
rise to an anomalous critical exponent for the fermion propagator.
This situation is quite analogous to the superconducting
instability. Actually, there is another Aslamasov-Larkin vertex
correction in the particle-particle channel, competing with the
$2k_{F}$ particle-hole instability. Inserting these
Aslamasov-Larkin vertex corrections into the shaded regions of
Fig. 1, one can construct self-consistent equations for fermion
and boson self-energies.} \label{fig3}
\end{figure}

This discussion reminds us of the conserving self-consistent
t-matrix approximation (CTMA) for the single impurity problem
\cite{LW_Impurity}, introduced to overcome the failure of the
non-crossing approximation (NCA) in the exactly screened case,
i.e., the absence of the strong coupling fixed point below the
Kondo temperature. Here, the NCA is analogous to the Eliashberg
approximation while the CTMA is parallel with the self-consistent
$2k_{F}$ treatment. Although the CTMA does not change the critical
exponents of the NCA in the over-screened case, it cures several
problems associated with thermodynamics \cite{LW_Impurity}. In the
present situation the self-consistent $2k_{F}$ treatment may
change the critical exponents of the Eliashberg approximation. It
will be really interesting to investigate this type of diagrams
near future.

\section{Summary}

In this study we solved three coupled integral equations for boson
and fermion self-energies with vertex corrections. The key point
is to make an ansatz for the vertex function in terms of the
fermion Green's function based on the Ward identity. Resorting to
this ansatz, we find a set of full self-consistent solutions in
the ladder approximation for the vertex correction, where both
boson and fermion self-energy corrections do not have any
modification, compared with the Eliashberg approximation. This
implies that if the anomalous exponent arises in the fermion
self-energy, it may result from the class of Aslamasov-Larkin
diagrams, not taken into account in our study. This remains as an
important future work.

K.-S. Kim appreciate an inspiring lecture of J. McGreevy in the
string spring school in ICTP (2010), motivating us to investigate
the present problem. K.-S. Kim was supported by the National
Research Foundation of Korea (NRF) grant funded by the Korea
government (MEST) (No. 2010-0074542).

\appendix

\section{Self-consistency between the Ward identity and the
ladder approximation for the vertex function in the low energy
limit}

In appendix A we show that the ladder vertex in Eq. (5) satisfies
the Ward identity with the ansatz Eq. (8). In other words, the
ansatz Eq. (8) is shown to be self-consistent at least in the
ladder approximation.

Inserting Eq. (8) into the right hand side (R.H.S) in Eq. (5), we
obtain \bqa && \mbox{R.H.S.} - 1 = - \frac{e^{2}}{N} \int \frac{d
l_{0}}{2\pi} \int\frac{d^{2} l}{(2\pi)^{2}} D(l_0,l) \nn &&
\frac{G_{\sigma}(k_0-l_0,k-l) - G_{\sigma}(k_0+q_0-l_0,k+q-l)
}{g_{\sigma}^{-1}(k_0+q_0-l_0,k+q-l) -
g_{\sigma}^{-1}(k_0-l_0,k-l)} . \nn \eqa Performing the $l_{x}$
integration, we obtain \bqa && \mbox{R.H.S.} - 1 = \frac{i
e^{2}}{2N} \int_{-\infty}^{\infty} \frac{d l_{0}}{2\pi}
\int_{-\infty}^{\infty} \frac{d l_y}{2 \pi} \frac{1}{\gamma
\frac{|l_{0}|}{|l_{y}|} + |l_{y}|^{z-1} } \nn &&
\frac{\mbox{sgn}(k_0-l_0) - \mbox{sgn}(k_0+q_0-l_0)}{i\eta q_{0} +
q_{x} + 2(k_y - l_y) q_y + q_{y}^{2}}  . \eqa Integrating over the
frequency $l_{0}$, we obtain \bqa && \mbox{R.H.S.} - 1 = \frac{i
e^{2}}{4 \pi \gamma N} \int_{0}^{\infty} \frac{d l_y}{2 \pi}
\frac{1}{2q_{y}} \Bigl\{ \ln \Bigl( \gamma |k_0+q_0| + l_{y}^{z}
\Bigr) \nn && - \ln \Bigl( \gamma |k_0| + l_{y}^{z}  \Bigr)
\Bigr\} \Bigl( \frac{ - i\eta q_{0} - q_{x} - 2k_y q_y - q_{y}^{2}
}{i\eta q_{0} + q_{x} + 2(k_y + l_y) q_y + q_{y}^{2}} \nn && +
\frac{ i\eta q_{0} + q_{x} + 2 k_y q_y + q_{y}^{2} }{i\eta q_{0} +
q_{x} + 2(k_y - l_y) q_y + q_{y}^{2}} \Bigr) \equiv \mathbf{A} +
\mathbf{B} . \eqa The first $\ln$ contribution can be approximated
as follows \bqa && \mathbf{A} \approx \frac{i e^{2}}{4 \pi \gamma
N} \int_{0}^{(\gamma |k_0+q_0|)^{1/z}} \frac{d l_y}{2 \pi}
\frac{1}{2q_{y}} \nn && \{ \ln ( \gamma |k_0+q_0| ) \} \Bigl(
\frac{ - i\eta q_{0} - q_{x} - 2k_y q_y - q_{y}^{2} }{i\eta q_{0}
+ q_{x} + 2(k_y + l_y) q_y + q_{y}^{2}} \nn && + \frac{ i\eta
q_{0} + q_{x} + 2 k_y q_y + q_{y}^{2} }{i\eta q_{0} + q_{x} +
2(k_y - l_y) q_y + q_{y}^{2}} \Bigr) \nn && + \frac{i e^{2}}{4 \pi
\gamma N} \int_{(\gamma |k_0+q_0|)^{1/z}}^{\Lambda} \frac{d l_y}{2
\pi} \frac{1}{2q_{y}} \nn && \{ \ln l_{y}^{z} \} \Bigl( \frac{ -
i\eta q_{0} - q_{x} - 2k_y q_y - q_{y}^{2} }{i\eta q_{0} + q_{x} +
2(k_y + l_y) q_y + q_{y}^{2}} \nn && + \frac{ i\eta q_{0} + q_{x}
+ 2 k_y q_y + q_{y}^{2} }{i\eta q_{0} + q_{x} + 2(k_y - l_y) q_y +
q_{y}^{2}} \Bigr) . \eqa

Expanding the first sector in $l_{y}$, we obtain \bqa &&
\mathbf{A} \approx \frac{i e^{2}}{4 \pi \gamma N}
\int_{0}^{(\gamma |k_0+q_0|)^{1/z}} \frac{d l_y}{2 \pi} \nn && \{
\ln ( \gamma |k_0+q_0| ) \} \frac{ 2l_{y}}{i\eta q_{0} + q_{x} + 2
k_y q_y + q_{y}^{2}}  \nn && + \frac{i e^{2}}{4 \pi \gamma N}
\int_{(\gamma |k_0+q_0|)^{1/z}}^{\Lambda} \frac{d l_y}{2 \pi}
\frac{1}{2q_{y}} \nn && \{ \ln l_{y}^{z} \} \Bigl( \frac{ - i\eta
q_{0} - q_{x} - 2k_y q_y - q_{y}^{2} }{i\eta q_{0} + q_{x} + 2(k_y
+ l_y) q_y + q_{y}^{2}} \nn && + \frac{ i\eta q_{0} + q_{x} + 2
k_y q_y + q_{y}^{2} }{i\eta q_{0} + q_{x} + 2(k_y - l_y) q_y +
q_{y}^{2}} \Bigr) . \eqa One will realize that the first term is
associated with the self-energy in the Eliashberg approximation.
Evaluating the $\mathbf{B}$ term in the same way as $\mathbf{A}$
and gathering both $\mathbf{A}$ and $\mathbf{B}$, we reach the
final expression \bqa && \Lambda(k_0+q_0,k+q;k_0,k) - 1 =
\mathbf{A} + \mathbf{B} \nn && = - \frac{\Sigma(k_0+q_0) -
\Sigma(k_0)}{g_{\sigma}^{-1}(k_0+q_0,k+q) -
g_{\sigma}^{-1}(k_0,k)} \nn && + \mathcal{F}(k_0+q_0,k+q;k_0,k) ,
\eqa where \bqa && \mathcal{F}(k_0+q_0,k+q;k_0,k) \equiv \frac{i
e^{2}}{4 \pi \gamma N} \int_{(\gamma |k_0+q_0|)^{1/z}}^{\Lambda}
\frac{d l_y}{2 \pi} \frac{1}{2q_{y}} \nn && \{ \ln l_{y}^{z} \}
\Bigl( \frac{ - i\eta q_{0} - q_{x} - 2k_y q_y - q_{y}^{2} }{i\eta
q_{0} + q_{x} + 2(k_y + l_y) q_y + q_{y}^{2}} \nn && + \frac{
i\eta q_{0} + q_{x} + 2 k_y q_y + q_{y}^{2} }{i\eta q_{0} + q_{x}
+ 2(k_y - l_y) q_y + q_{y}^{2}} \Bigr) \nn && - \frac{i e^{2}}{4
\pi \gamma N} \int_{(\gamma |k_0|)^{1/z}}^{\Lambda} \frac{d l_y}{2
\pi} \frac{1}{2q_{y}} \nn && \{ \ln l_{y}^{z} \} \Bigl( \frac{ -
i\eta q_{0} - q_{x} - 2k_y q_y - q_{y}^{2} }{i\eta q_{0} + q_{x} +
2(k_y + l_y) q_y + q_{y}^{2}} \nn && + \frac{ i\eta q_{0} + q_{x}
+ 2 k_y q_y + q_{y}^{2} }{i\eta q_{0} + q_{x} + 2(k_y - l_y) q_y +
q_{y}^{2}} \Bigr) . \eqa It is not difficult to observe that
$\mathcal{F}(k_0+q_0,k+q;k_0,k)$ is irrelevant in the low energy
limit due to the frequency and momentum dependence in the
numerator, giving rise to higher order corrections to the fermion
self-energy. We conclude that Eq. (8) is asymptotically correct in
the low energy limit.

\section{Evaluation of $\Sigma_{2}(k_0)$}

In appendix B we evaluate $\Sigma_{2}(k_{0})$. Performing momentum
and frequency integrals, we obtain \bqa && \Sigma_{2}(k_0) =
\frac{e^{2}}{N} \int \frac{d q_{0}}{2\pi} \int \frac{d^{2}
q}{(2\pi)^{2}} \frac{1}{i\eta q_{0} + q_{x} + 2k_y q_y + q_{y}^{2}
} \nn && \frac{i \eta k_0 + k_x + k_y^{2} - \Sigma(k_0)}{i \eta
(k_0+q_0) + (k_x + q_x) + (k_y+q_y)^{2} - \Sigma(k_0+q_0)} \nn &&
\frac{1}{\gamma \frac{|q_{0}|}{|q_{y}|} + |q_{y}|^{z-1} } \nn && =
\frac{e^{2}}{N} \int \frac{d q_{0}}{2\pi} \int \frac{d
q_{y}}{2\pi} \int \frac{d q_{x}}{2\pi} \frac{1}{\gamma
\frac{|q_{0}|}{|q_{y}|} + |q_{y}|^{z-1} } \nn && \frac{i \eta k_0
+ k_x + k_y^{2} - \Sigma(k_0)}{i \eta k_{0} + k_{x} + k_{y}^{2} -
\Sigma(k_0+q_0)} \Bigl\{ \frac{1}{(i\eta q_{0} + q_{x} + 2k_y q_y
+ q_{y}^{2})} \nn && - \frac{1}{i \eta (k_0+q_0) + (k_x + q_x) +
(k_y+q_y)^{2} - \Sigma(k_0+q_0)} \Bigr\}
\nn && = - \frac{i e^{2}}{2 \pi \gamma^{1-2/z} N} \Bigl\{
\int_{0}^{\Lambda} d y \frac{y}{1 + y^{z} } \Bigr\} \nn && \int
\frac{d q_{0}}{2\pi} \frac{i \eta k_0 + k_x + k_y^{2} -
\Sigma(k_0)}{i \eta k_{0} + k_{x} + k_{y}^{2} - \Sigma(k_0+q_0)}
\nn && \frac{\mbox{sgn}(q_0) - \mbox{sgn}(k_0+q_0)}{q_{0}^{1-2/z}}
\approx - i \frac{\lambda }{N} \mbox{sgn}(k_0) |k_0|^{2/z} .  \eqa
The vertex correction does not change the scaling for frequency in
the ladder approximation.

\end{document}